\documentclass[a4paper]{article}

\usepackage{INTERSPEECH2021}
\usepackage{subfigure}
\usepackage{multicol}
\usepackage{enumitem}
\usepackage{url}

\title{WSRGlow: A Glow-based Waveform Generative Model for Audio Super-Resolution}
\name{Kexun Zhang$^1$, Yi Ren$^1$, Changliang Xu$^2$, Zhou Zhao$^{1}$}
\address{
  $^1$Zhejiang University\\
  $^2$State Key Laboratory of Media Convergence Production Technology and Systems}
\email{\{kexunz, rayeren, zhaozhou\}@zju.edu.cn, xu@shuwen.com}

\begin{document}

\maketitle
\begin{abstract}

Audio super-resolution is the task of constructing a high-resolution (HR) audio from a low-resolution (LR) audio by adding the missing band. Previous methods based on convolutional neural networks and mean squared error training objective have relatively low performance, while adversarial generative models are difficult to train and tune. Recently, normalizing flow has attracted a lot of attention for its high performance, simple training and fast inference. In this paper, we propose WSRGlow, a Glow-based waveform generative model to perform audio super-resolution. Specifically, 1) we integrate WaveNet and Glow to directly maximize the exact likelihood of the target HR audio conditioned on LR information; and 2) to exploit the audio information from low-resolution audio, we propose an LR audio encoder and an STFT encoder, which encode the LR information from the time domain and frequency domain respectively. The experimental results show that the proposed model is easier to train and outperforms the previous works in terms of both objective and perceptual quality. WSRGlow is also the first model to produce 48kHz waveforms from 12kHz LR audio. Audio samples are available at \url{https://zkx06111.github.io/wsrglow/}.

\end{abstract}

\section{Introduction}

Audio super-resolution is the task of constructing a high-resolution (HR) audio from a low-resolution (LR) audio by adding the missing band in a coherent manner. Formally, given a LR sequence of audio samples $x_l=(x_{1/R_l}, \dots, x_{R_lT/R_l})$, we wish to synthesize a high-resolution audio signal $x_h=(x_{1/R_h}, \dots, x_{R_hT/R_h})$, where $R_l$ and $R_h$ are the sampling rates of the low and high-resolution signals. Audio super-resolution can be used to achieve better speech quality with a limited sampling rate in telecommunications. It can also improve the audio quality of ancient recordings and music, and facilitate TTS~\cite{ren2019fastspeech,ren2020fastspeech,ren2019almost,xu2020lrspeech,chen2020multispeech} and singing voice synthesis~\cite{ren2020deepsinger,liu2021diffsinger} models to generate higher-quality voices.
  
Previous approaches to this task include Gaussian mixture models \cite{cheng1994gaussian,pulakka2011gaussian}, linear predictive coding \cite{bradbury2000linear} and neural networks \cite{Li2015DNNbasedSB}. Learning-based methods perform better in this context because they can capture sophisticated domain-specific information. Convolutional networks \cite{kuleshov2017,Lim2018TimeFrequencyNF} and generative adversarial networks (GAN)  \cite{kim2018adversarial,eskimez2019,Hu2020} have shown to greatly improve the quality of the synthesized high-resolution audio. However, GANs are known to be hard to train and produce outputs with significant artifacts. Recently, the success of normalizing flow models has been used on many generative tasks due to its ability to optimize the log-likelihood exactly and parallelizability of both training and synthesis. Generative glow is capable of efficient realistic-looking synthesis and manipulation of large images \cite{kingma2018glow, CFlow, Kondo2019FlowbasedIT}. Normalizing flow conditioned on low-resolution images \cite{lugmayr2020srflow} has also improved the performance of image super-resolution, which is similar to speech super-resolution in many ways.
  
Considering that audio super-resolution is a one-to-many mapping (ill-posed) problem since multiple possible HR audio samples correspond to a given LR one, in this paper, we explore the potential of normalizing flow on modeling the conditional distribution of HR audio samples given an LR one. Specifically, we propose WSRGlow, a novel model for audio super-resolution based on generative flows. WSRGlow generates HR audio by sampling from a simple zero mean spherical Gaussian distribution. The sample taken from the simple distribution is then put through several invertible layers and transformed to the desired output audio. The layers of our model are similar to those in WaveGlow \cite{prenger2019waveglow} and WaveNet \cite{oord2016wavenet}. During training, HR audio is fed to the model as input to obtain the corresponding sample points in the simple distribution, so that we can train our model by directly maximizing the likelihood. During inference, WSRGlow takes a sample from the simple distribution and passes it through several invertible layers to compute the desired output audio. 
  
To better extract the valuable information in LR audio, we propose two encoders: an LR audio encoder and an STFT encoder, which encode LR information from the time and frequency domains, respectively. The LR audio encoder applies $\mu$-law transformation to raw low-resolution waveform \cite{mulaw}, quantizes each sample to mu-law code and maps the code sequence to embedding vectors. However, the LR audio encoder only provides a time-domain perspective and it is challenging for LR audio encoder to exploit several characteristics of the signal, including cyclic behavior and long-range dependence, due to the limited receptive field and model capacity. Therefore, we further introduce the frequency domain encoder called STFT encoder, which first applies short-time Fourier transform (STFT) to the waveform and then encodes the phase and magnitude spectrogram, both of which are widely used in modern speech applications and methods~\cite{Hu2020, pwn4sr}. The contributions of this work are summarized as follows:
\begin{itemize}[leftmargin=*]
    \item To the best of our knowledge, WSRGlow is the first model to use normalizing flow structure for audio super-resolution task and proves the potential of normalizing flow for audio super-resolution.
    \item We propose an LR audio encoder and an STFT encoder to exploit LR information from both time domain and frequency domain. 
    \item Our experiments on VCTK \cite{vctk} dataset show that 1) the proposed model is easier to train and outperforms the previous works both objectively and perceptually; and 2) STFT encoder improves the model performance significantly\footnote{Audio samples are available at \url{https://zkx06111.github.io/wsrglow/}. The codes for this paper is available at \url{https://github.com/zkx06111/WSRGlow}.} and makes the convergence much faster. 
\end{itemize}

\section{Model Architecture}

\begin{figure}[!h]
  \centering
  \includegraphics[width=0.9\linewidth]{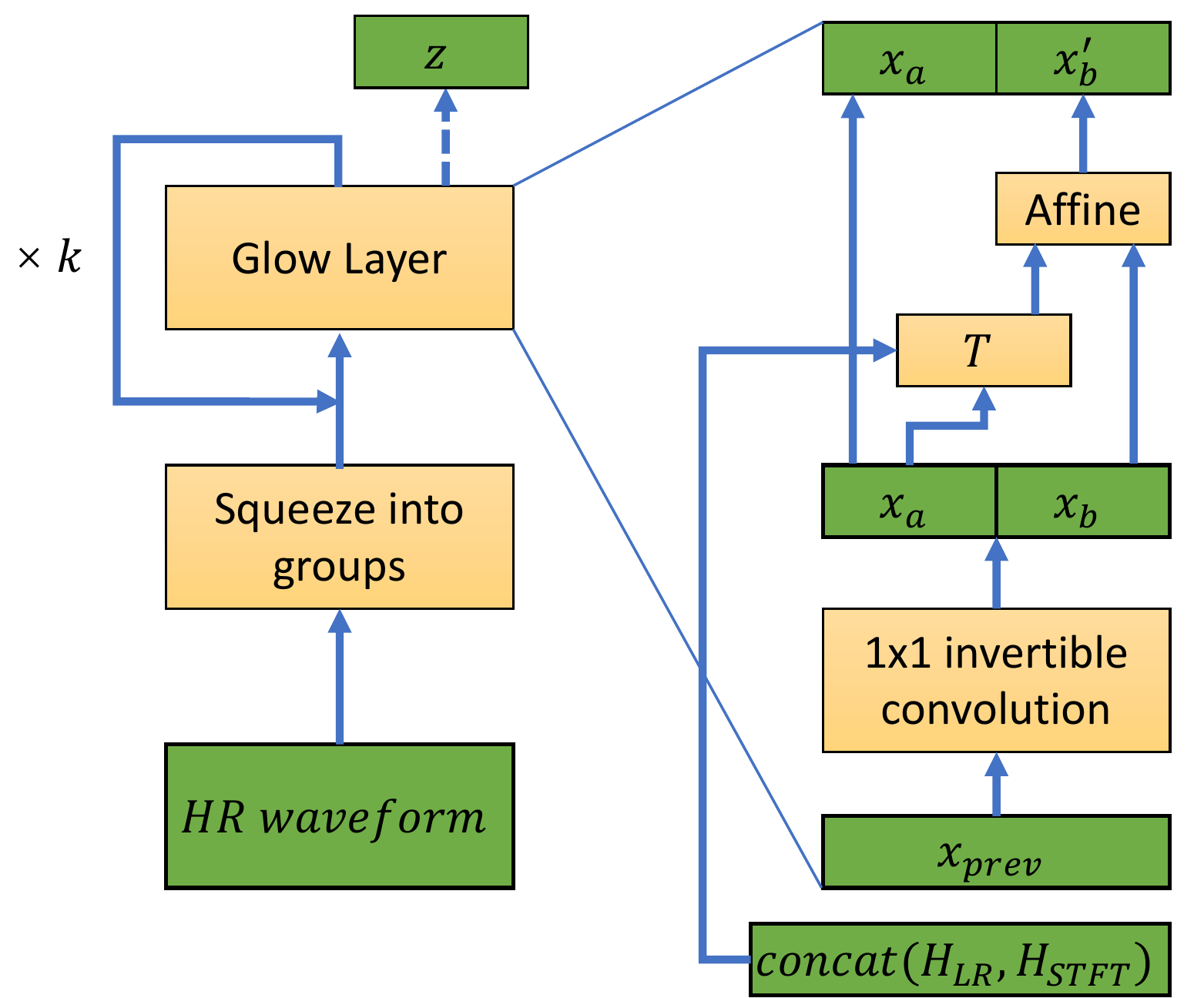}
  \caption{Overall architecture. During training, the model runs along the direction of the arrows, while the model runs inversely during inference (except the $T$ transformation). The two conditional encoders' outputs are concatenated to be the condition for transformation $T$.}
  \label{fig:arch}
\end{figure}

Audio super-resolution is an ill-posed problem whose solution is not unique. Various possible solutions exist and form a distribution in the solution space. To capture the probabilistic nature of super-resolution audio, we employ normalizing flow to model the conditional distribution of HR audio given the LR audio. WSRGlow captures the bijection between a zero mean spherical Gaussian distribution and the probability space of audio waveform using invertible layers similar to Glow \cite{kingma2018glow}. To generate high-resolution audio, we encode the low-resolution audio using an LR audio encoder and an STFT encoder and then use the encoding as the condition for the glow layers. The architecture of WSRGlow is depicted in Figure \ref{fig:arch}. We describe the details of each component below.

\subsection{Glow Layers}

To generate high-resolution audio, we take a sample (${\bm z} \sim \mathcal{N}({\bm z};0,{\bm I})$) from a simple Gaussian distribution and passes it through several invertible layers $\bm f_{i}$ that transforms the simple distribution to the desired distribution $\bm x$ in the audio space:
\begin{equation}
{\bm x} = {\bm f}_0\circ{\bm f}_1\circ\dots \circ{\bm f}_k({\bm z}).\nonumber
\end{equation}
During training, high-resolution audio is regarded as a vector and squeezed into groups of 8, as in WaveGlow \cite{prenger2019waveglow}. This vector is then passed through the invertible layers and transformed into a simple distribution, so that we can optimize the negative loglikelihood directly. The invertibility of the WSRGlow layers ensure that we can calculate the loglikelihood $\log p_{\theta}(\bm x)$ directly using inverted mappings:
\begin{equation}
    \log p_{\theta}({\bm x}) = \log p_{\theta}({\bm z})+\sum_{i=1}^k \log |\det ({\bm J}({\bm f}_{i}^{-1}(x)))|,\nonumber
\end{equation}
\begin{equation}
    {\bm z} = \bm f_{k}^{-1} \circ \bm f_{k-1}^{-1}\circ \dots \circ \bm f_{0}^{-1}(x),\nonumber
\end{equation}
where $\bm J$ is the Jacobian of the inversed transformation $\bm f_{i}^{-1}$. The invertible layers have a similar structure to those in WaveGlow \cite{prenger2019waveglow}. Each of the WSRGlow layers consists of an invertible 1x1 convolution $\bm f_{conv}$ and an affine coupling layer $\bm f_{coupling}$. The detailed structure of these layers will be described below.

\subsubsection{Affine Coupling Layer}

We use an affine coupling layer $\bm f_{coupling}$ to make a WSRGlow layer invertible. The channels of the input tensor $\bm x$ is split into two halves $\bm x_a$ and $\bm x_b$. Half of the channels are passed through a transformation $T$ to produce additive term $\bm t$ and multiplicative term $\bm s$ for the other half.
\begin{equation}
    \bm x_a, \bm x_b = split(x),\nonumber
\end{equation}
\begin{equation}
    (\log \bm s, \bm t)=T(\bm x_a, concat(H_{LR}, H_{STFT})),\nonumber
\end{equation}
\begin{equation}
    \bm x_b'=\bm s\odot \bm x_b + \bm t,
\end{equation}
\begin{equation}
    \bm f^{-1}_{coupling}(\bm x)=concat(\bm x_a, \bm x_b'),\nonumber
\end{equation}
\begin{equation}
    \log |\det (\bm J(\bm f^{-1}_{coupling}(\bm x))|=\log |\bm s|.\nonumber
\end{equation}
Here $T$ can be any transformation that does not have to be invertible because the coupling layer keeps the invertibility of the network. When inverting the network, since $\bm x_a$ is passed without changing, we can obtain $\bm s$ and $\bm t$ by recomputing $T$ with $\bm x_a$. We can then solve equation (1) for $\bm x_b$ using $\bm s, \bm t, \bm x_b'$: $\bm x_b = \frac{\bm x_b - \bm t}{\bm s}$.

In our model, the transformation $T$ is similar to the structure of non-causal WaveNet \cite{oord2016wavenet} with dilated convolutions, gated-$\tanh$ activations and residual connections. The main difference between $T$ and non-causal WaveNet is that the condition parameter for $T$ is not a mel-spectrogram, but the encoded low-resolution audio signal $concat(H_{LR}, H_{STFT})$.

\subsubsection{1x1 Invertible Convolution}

Following Glow \cite{kingma2018glow}, we employ a 1x1 invertible convolution layer $\bm f_{conv}^{-1} = \bm W \bm x$ before every affine layer to enable two halves of the channels to modify the one another. The convolution kernel is initialized to be orthonormal to keep its invertibility. 

\subsection{Conditional Encoder}

\begin{figure}[ht!]
  \centering
  \subfigure[LR Audio Encoder]{
      \includegraphics[width=0.4\linewidth]{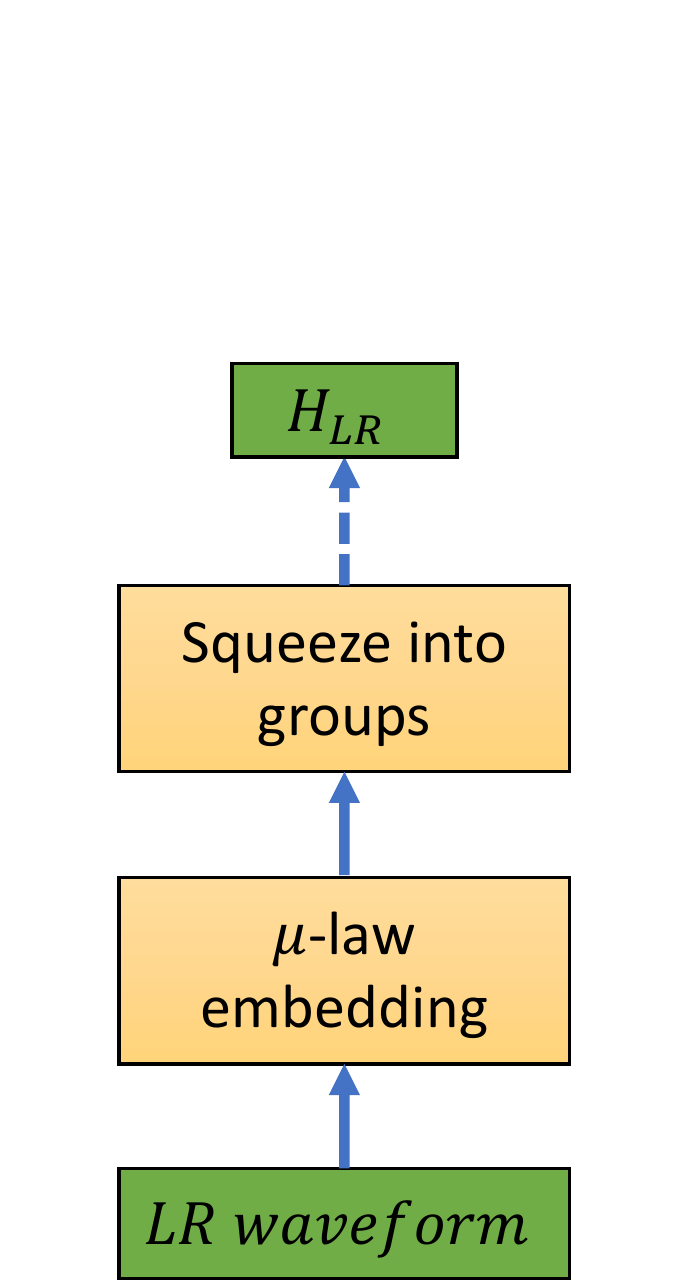}
  }
  \subfigure[STFT Encoder]{
      \includegraphics[width=0.4\linewidth]{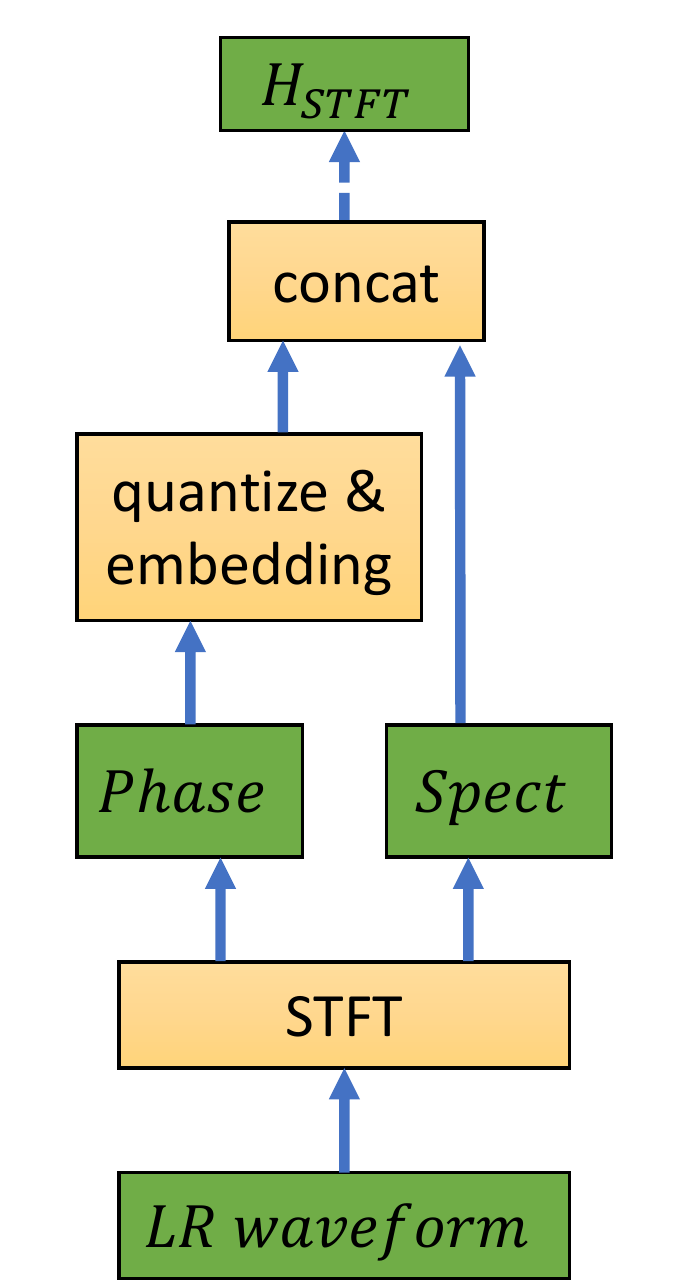}
  }
  \caption{Architecture of conditional encoders.}
  \label{fig:enc}
\end{figure}

We propose two encoders to extract information from LR audio for conditioning WSRGlow. One is the LR audio encoder which directly maps raw audio samples to a sequence of embedding vectors. The other is the STFT encoder which encodes LR audio from the frequency domain by encoding its magnitude and phase spectrogram. The size of the group for squeezing LR encoding is chosen to be the same as the time window of STFT, so that the encodings of the two encoders can be concatenated to encode LR audio from both time and frequency domains. The architecture of the two encoders is shown in Figure \ref{fig:enc}.

\subsubsection{LR Audio Encoder}

The LR audio encoder $Enc_{LR}$ maps every sample in the low-resolution signal to an embedding space. For a sample $x_t$, we first apply the $\mu$-law transformation as in WaveNet \cite{oord2016wavenet}, and then quantize the transformed sample to 256 possible values. We create an embedding vector for each of the 256 possible values. The sequence of embedding vectors is then squeezed into a sequence of groups of vectors $H_{LR}$ to match the dimensionality of $T$.

\subsubsection{STFT Encoder}

The STFT encoder $Enc_{STFT}$ encodes the frequency-domain information in LR audio, including magnitude and phase spectrograms. While magnitude spectrogram is a well-known representation, we also introduce phase spectrogram since it has been proven to be of much help in various speech tasks \cite{Paliwal2011TheIO} including speech enhancement, speech recognition and speech synthesis. Besides, it is also shown that the capacity of human perception due to phase is higher than expected \cite{Mahale2014PhaseII}.

The STFT encoder first applies the short-time Fourier transform (STFT) to the signal. The phases $\phi$ of each time window is quantized and then embedded to $E_{\phi}$:
\begin{equation}
    E_{\phi}(\bm x)=\text{Embedding}(\text{Quantize}(\text{imag}(\text{STFT}(\bm x))),\nonumber
\end{equation}
while the magnitude spectrogram $S$ is directly taken: $S(\bm x)=\text{real}(\text{STFT}(\bm x))$. The two are then concatenated along the channel axis to obtain the STFT encoding $H_{STFT}$: $H_{STFT}=concat(S(\bm x), E_{\phi}(\bm x))$.

\section{Experiments}
\subsection{Experimental Settings}
\subsubsection{Dataset}

We evaluate our model on the VCTK corpus \cite{vctk}, which contains 44 hours of data from 108 different speakers. We generate low-resolution audio signal using the \textit{resample} function with default parameters in Librosa \cite{librosa}. The target sampling rate of HR audio is 48kHz. We randomly split the dataset into training set and test set at utterance level so that the training set also contains the speakers in the test set. The audio file names used in test set are listed in the supplementary materials.

\subsubsection{Model Configurations}
Four methods for speech super-resolution are implemented for comparison, including two baselines and two proposed WSRGlow models with different LR encodings.

\begin{itemize}[leftmargin=*]
    \item AudioUNet: A deep convolutional model for audio super-resolution proposed in \cite{kuleshov2017}. We instantiated the model with 8 down-sampling blocks and 8 up-sampling blocks. We used the same parameters for each block as \cite{kuleshov2017} and our PyTorch version of AudioUNet is a reimplementation of the TensorFlow code provided by the authors\footnote{https://github.com/kuleshov/audio-super-res}.
    \item MUGAN: A GAN-based model for audio super-resolution proposed in \cite{eskimez2019,kim2018adversarial}. We use a similar structure in AudioUNet for the generator and the discriminator. We use MSE Loss for the generator and cross-entropy loss for the discriminator.
    \item WSRGlow w/o $Enc_{STFT}$: WSRGlow with only the encoding of LR audio encoder as the condition. We set the embedding dimension to be 256 and the number of glow layers to be 12. Audio encodings are squeezed into groups of 8. The transformation $T$ in the affine coupling layer has 8 layers of dilated convolutions with 256 channels.
    \item WSRGlow: WSRGlow with the concatenated encoding of both LR audio encoder and STFT encoder as the condition. WSRGlow follows the setting of WSRGlow w/o $Enc_{STFT}$. We set the embedding size of the phase spectrogram to be 50. The frame size for STFT is set to 8 so that the time dimensionality of phase embeddings and waveform embeddings can be the same.
\end{itemize}

\subsubsection{Training and Evaluation}

For all models, we train them on batches of 12 audio waveforms with the maximum number of HR samples being 8192. We train each model for 100k iterations. For AudioUNet, WSRGlow w/o $Enc_{STFT}$ and WSRGlow, we use the ADAM~\cite{ADAM} optimizer with learning rate being 1e-4, $\beta_1=0.9$, $\beta_2=0.98$. For MUGAN, We set the coefficient for adversarial loss to be 0.001 and start adding the adversarial term to the loss for generator after 40k steps.

Following previous works \cite{kuleshov2017,kim2018adversarial} on speech super-resolution, we evaluate our experiment results using both objective metrics and perceptive surveys.

We use the Signal to Noise Ratio (SNR) and Log-spectral distance (LSD) \cite{LSD} to measure the quality of generated audio samples by comparing them to the actual high-resolution audio. The SNR is defined as $\text{SNR}(x, y)=10\log \frac{||y||_2^2}{||x-y||_2^2}$,
where $x$ is the reference signal and $y$ is an approximation. The LSD measures the reconstruction quality of individual frequencies, which is defined as $
    \text{LSD}(x, y)=\frac{1}{L}\sum_{l=1}^L\sqrt{\frac{1}{K}\sum_{k=1}^K \left(X(l,k)-\hat X(l,k)\right)^2}$,
where $X$ and $\hat X$ are the log-spectral power magnitudes of $y$ and $x$. They are defined as $X=\log |S|^2$ where $S$ is the short-time Fourier transform of the signal. $l$ is the number of frames and $k$ is the number of frequencies. We use the evaluation setting in \cite{kuleshov2017} where the length of frames is 2048.

Following MUGAN \cite{kim2018adversarial}, we perform a randomized, single-blinded A/B test with 10 participants to evaluate the performance of WSRGlow with real listeners. The test presents 10 pairs of audio clips generated by WSRGlow with STFT encoding and the best baseline model MUGAN, and asks participants to select a preferred clip, or ``Neutral".

\begin{figure*}[!h]
  \centering
  \subfigure[Ground Truth]{
      \includegraphics[width=0.23\linewidth]{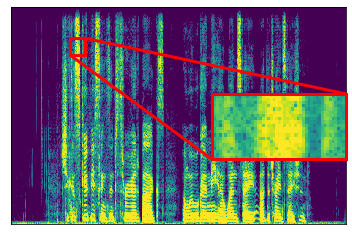}
  }
  \subfigure[AudioUNet]{
      \includegraphics[width=0.23\linewidth]{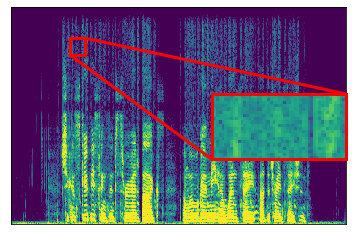}
  }
  \subfigure[MUGAN]{
      \includegraphics[width=0.23\linewidth]{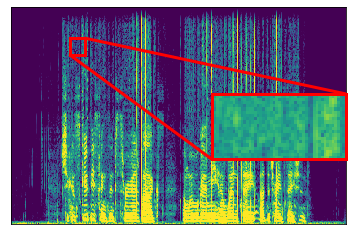}
  }
  \subfigure[WSRGlow]{
      \includegraphics[width=0.23\linewidth]{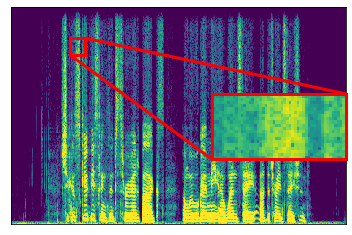}
  }
  \vspace{-2mm}
  \caption{Spectrogram visualizations of \textit{p225\_005.wav}.}
  \vspace{-2mm}
  \label{fig:spect}
\end{figure*}

\subsection{Results and Analyses}
\subsubsection{Subjective Evaluation}

We conduct the preference test to compare the perceptual quality of the audio generated by WSRGlow and one of the baseline methods including AudioUNet and MUGAN. The results are shown in Table \ref{tab:abtest}. From the table, we can see that listeners prefer more audio samples generated by WSRGlow over the baseline methods AudioUNet and MUGAN. We observe that the audio generated by MUGAN contains artifacts that do not sound like audio voiced by human beings. While this kind of artifacts does not exist in audio generated by WSRGlow, indicating that our method can reconstruct the natural and high-quality high-frequency audio.

We also visualize the spectrograms of ground truth audio and synthesized 4x super-resolution samples of AudioUNet, MUGAN and WSRGlow in Figure \ref{fig:spect}. We observe that the reconstructed part of WSRGlow is much closer to the ground truth than the baseline models, which is consistent with the preference test.

\begin{table}[h]
\centering
\footnotesize
\vspace{-2mm}
\caption{The preference scores of baselines and WSRGlow.}
\begin{tabular}{c|c|c|c}
\toprule
Baseline & Baseline wins & Neutral & WSRGlow wins \\ \midrule
AudioUNet &   17\%       & 16\% &          67\% \\ 
MUGAN     &    18\%      & 20\% &          62\%  \\
\bottomrule
\end{tabular}
\vspace{-3mm}
\label{tab:abtest}
\end{table}

\subsubsection{Objective Evaluation}

We also evaluate the objective metrics including SNR and LSD for WSRGlow and baseline methods. The results are shown in Table \ref{tab:objmetrics}. From row 1, 2 and 4, we can see that WSRGlow outperforms the baselines and WSRGlow w/o $Enc_{STFT}$ significantly, which further demonstrates the superiority of our method. Besides, we remove the STFT encoder $Enc_{STFT}$ in WSRGlow and compare it with the original WSRGlow. The results are shown in row 3 (WSRGlow w/o $Enc_{STFT}$). We can see that WSRGlow w/o $Enc_{STFT}$ outperforms the baseline models in terms of SNR, but performs poorly in LSD. This indicates that WSRGlow w/o $Enc_{STFT}$ cannot synthesize the high-frequency part of the signal well. Because SNR is calculated using sample-wise distance, while LSD is calculated using the spectrogram. A low LSD score also suggests unsatisfactory perceptual quality since the LSD metric has better correlation with perceptual quality compared to SNR \cite{kuleshov2017,LSD}. Besides, we observe that WSRGlow performs much better in LSD score than WSRGlow w/o $Enc_{STFT}$. This indicates that encoding the LR audio from frequency domain is of great help in synthesizing the high-frequency part of the audio.

\begin{table}[h]
\footnotesize
\centering
\vspace{-2mm}
\caption{Objective comparison with other methods.}
\begin{tabular}{c|rr|rr}
\toprule
\multicolumn{1}{l}{} & \multicolumn{2}{|c|}{2x SR}                         & \multicolumn{2}{c}{4x SR}                        \\
\multicolumn{1}{r|}{} & \multicolumn{1}{l}{SNR$\uparrow$} & \multicolumn{1}{l|}{LSD$\downarrow$} & \multicolumn{1}{l}{SNR$\uparrow$} & \multicolumn{1}{l}{LSD$\downarrow$} \\ \midrule
AudioUNet            & 22.68                   & 1.01                     & 17.15                   & 2.24                    \\
MUGAN                & 24.81                   & 0.90                     & 16.87                   & 2.12                    \\
WSRGlow w/o $Enc_{STFT}$ & 25.91                   & 1.41                     & 17.45                   & 2.76                    \\
WSRGlow         & \textbf{30.84}          & \textbf{0.68}            & \textbf{18.38}          & \textbf{1.64}           \\ \bottomrule
\end{tabular}
\vspace{-3mm}
\label{tab:objmetrics}
\end{table}

\subsubsection{Ablation Study}
To verify the effectiveness of our proposed components and analyze our methods, we conduct ablation study in three aspects:

\textbf{Convergence speed:} We train WSRGlow on 4x super-resolution task for only 50k iterations (which is half of the iterations run for the baselines) to show that WSRGlow is easy to train. After 50k iterations, WSRGlow achieves $\text{SNR}=17.53, \text{LSD}=1.80$ on the test set, which is already better than the baseline models. This proves that WSRGlow converges at high speed.

\textbf{Necessity of phase and magnitude inputs:} The comparison between WSRGlow and WSRGlow w/o $Enc_{STFT}$ in Table \ref{tab:objmetrics} already shows that STFT information (phase and magnitude) is crucial in WSRGlow. To examine whether one of the two can be omitted, we train two partial WSRGlow models for 60k iterations. We observe that WSRGlow without phase embeddings results in $SNR=12.91, LSD=2.35$ while WSRGlow without magnitude embeddings results in $SNR=5.68, LSD=2.32$. Both results are poor compared to even the baseline models. This proves that both $Enc_{LR}$ and $Enc_{STFT}$ are effective and necessary.

\textbf{Temperature in sampling:} We sample the high-frequency audio results using the same WSRGlow model with different temperatures (variance of Gaussian) $T=0.5, 0.666, 0.8, 1.0, 1.2, 1.5$ to explore the effect on audio quality. We observe that as the temperature becomes lower, SNR score tends to increase while LSD score tends to decrease. Perceptually, we find that the speech quality is the best when $T=1.0$.\footnote{Due to the limitation of space, we put audio samples and objective evaluation for different temperatures in the supplementary materials.}

\section{Conclusion}

In this paper, we proposed WSRGlow, a glow-based generative model for audio super-resolution: 1) we use normalizing flow to address the one-to-many problem by modeling the conditional distribution of HR audio; and 2) we proposed an LR encoder and an STFT encoder to exploit the information in LR audio from both time domain and frequency domain, significantly improving model's ability to synthesize the high-frequency part. Our experiments demonstrated that WSRGlow outperforms previous models in both objective and subjective metrics and converges quickly. The components were verified to be necessary and effective through ablation study.

In the future, we may develop our model to address more generalized audio super-resolution tasks, such as speech in heterogeneous environments and music recordings. We may also model the conditional distribution of HR spectrogram instead of HR waveform.

\bibliographystyle{IEEEtran}

\bibliography{mybib}

\end{document}